# Strain and spin orbit coupling effects on electronic and optical properties of 2D CX/graphene (X = S, Se, Te) vdW heterostructure for solar energy harvesting

Amit K. Bhojani,[1] Hardik L. Kagdada[2] and Dheeraj K. Singh[1, *]

[1]Department of Basic Sciences, Institute of Infrastructure, Technology, Research And Management (IITRAM), Ahmedabad 380026, India

[2]Department of Mechanical Engineering, Indian Institute of Technology Bombay, Mumbai 400076, India

**\*Corresponding author:** dheerajsingh84@gmail.com; dheerajsingh@iitram.ac.in

**Abstract:** Vertically stacked two-dimensional (2D) graphene-based van der Waals (vdW) heterostructures have emerged as the technological materials for electronic and optoelectronic device applications. In this regard, for the first time, we systematically predicted the electronic and optical properties of CX/G (X = S, Se and Te; G = graphene) heterostructures under biaxial strain and spin orbit coupling (SOC) by first-principles calculations. Strain is induced by applying mechanical stress to the heterostructures, while SOC arises due to the interaction between the electron spin and its orbital motion. The electronic property calculations reveal that all three heterostructures exhibit indirect semiconducting nature with a narrow bandgap of 0.47-0.62 $eV$ and remain indirect under compressive and tensile strains. Strong band splitting of 78.4 $meV$ has been observed in the conduction band edge of CTe/G heterostructure in the presence of SOC due to the lack of an inversion center attributed to the large hole effective mass. Under compressive strain, the p-type of Schottky contact of CX/G heterostructures is converted into p-type Ohmic contact because of nearly negligible Schottky barrier height. Further, the optical property assessment reveals red and blue shifts in the absorption peak of CX/G heterostructures with regard to tensile and compressive strains, respectively. Despite this, the CTe/G heterostructure achieves a remarkable high $\eta$ of 24.53% in the strain-free case whereas, it reaches to 28.31% with 4% compressive strain, demonstrating the potential for solar energy conversion device applications. Our findings suggest that CX/G heterostructures could be promising candidates for high-performance optoelectronic devices.

# 1. Introduction

The discovery of graphene in 2004 by Geim and co-workers via mechanical exfoliation leads to the new era of two-dimensional (2D) materials with exceptionally high carrier mobility and mechanical strength, and high thermal conductivity [27]. Despite this novel features, the zero electronic bandgap of graphene due to Dirac point at band edges limits its practical applications as semiconducting material. Further, the light absorption capability of monolayer graphene is found to ~2.3%, which is undesirable for their high-performance in optoelectronic devices [12]. Although, the discovery of graphene leads to the search for variety of 2D materials including hexagonal boron nitride, transition metal dichalcogenides, MXenes, group-III and IV-based pnictides, etc. have been widely explored experimentally as well as theoretically due to their superior physical and chemical properties for nanoscale applications [25,33,49]. It is a well-known fact that the properties of the 2D materials are way more different than that of bulk counterparts.

There are numerous possibilities to stack the various 2D materials in one vertical line, which might be held together vertically by the van der Waals (vdW) forces, while in-plane stability could be achieved through covalent interactions. When 2D materials are stacked on top of each other, the redistribution of charges and the structural changes induces and can be controlled through the elements present in the heterostructures [28]. vdW Heterostructures offers plethora of opportunities for alternative solutions of the graphene limitations [11,22,44]. For example, the stacking of graphene on the surface of other 2D semiconducting materials form graphene-based vdW heterostructures, which effectively tunes the electronic properties of graphene because of unique interfaces and electron coupling effect [31]. The graphene-based heterostructures forge graphene as a suitable candidate for electronic devices by introducing the bandgap of about a few tens at Dirac K point [31]. Further, it is possible to maintain the carrier mobility as higher as in perfect graphene in these heterostructures [5,30]. Moreover, the formation of Schottky junctions in graphene-based vdW heterostructures enhances the carrier efficiencies between the graphene and semiconducting materials [12]. These finding indicate that graphene-based vdW heterostructures exhibit a wide range of applications for future nanoelectronic and optoelectronic devices.

In the past decade, a variety of graphene-based vdW heterostructures have been successfully synthesized and theoretically investigated, such as graphene/h-BN [42], graphene/group-IV chalcogenides, graphene/transition metal dichalcogenides [16], graphene/MXenes [20], etc. for their diverse applications. while DFT based investigation of Hone et al. on the graphene and hexagonal boron nitride (h-BN) heterostructure reveals that the reduction of the charge scattering site is responsible for high carrier mobility. Further confirming that the electron coupling interaction between the graphene and h-BN shows the major impact on the physical properties of heterostructures [7]. Another investigation on graphene and transition metal dichalcogenides (TMDs) heterostructures reveals that the bandgap of graphene is opened due to the overlapping of the π-band of graphene and out-of-plane orbital components of TMDs electronic band structure [32]. Further, the graphene and group-IV-based monochalcogenide heterostructures also show their potential for the next generation nanoelectronic and optoelectronic devices [5,41].

Moreover, the strain impact on 2D materials have already demonstrated their relevance to control the electronic properties (direct-indirect band gap and semiconductor-metallic transition), catalytic activities, and energy storage capacity [6,10,43,46]. Further, the optical properties of 2D materials can also be modified using the external strain as a tunable function i.e., strain induces the differences in optical leap behaviors [20]. Similarly, in 2D heterostructures, several reports demonstrate the effect of external strain on the physical properties such as, strain effectively modulates the variation in band arrangement from type-I to type-II or III and it adjusts the Schottky barrier height of the heterostructures [5,36,41]. Sun et al. disclosed that unstrained graphene/GaN heterostructures possessed n-type of Schottky contact while it can be converted into p-type in the presence of external vertical strain, suggesting application for next-generation Schottky devices [36]. Similarly, a recent study on graphene/GeS vdW heterostructure demonstrates that normal strain tunes the p-type Schottky barrier height with improved structural rigidity and anisotropy of heterostructures compared to monolayer GeS, rendering their potential for high-performance nanoelectronic devices [5]. Recently, by using DFT calculations, Tuan et al. show that the graphene/WSeTe heterostructures improve optical absorption range in both visible and UV regions compared to their respective monolayers, additionally, the obtained excellent carrier mobility and bandgap of about 10 meV at Dirac K point can be tuned using vertical strain or electric field, demonstrating their high-speed electronic device applications [41]. Furthermore, strain engineering is also utilized to control the spin orbit coupling (SOC) effects, which provides possibilities of manipulating the electron spin in many 2D semiconducting heterostructures [8,35,39,47].

Despite having excellent reports on various graphene-based vdW heterostructures, the literature lacks a complete analysis of the electronic and optical properties of graphene and carbon-based monochalcogenides heterostructures. The motivation came from outstanding demonstration of carbon-based monochalcogenides for optoelectronic device applications [1,18,37]. Therefore, the present work reports for the first time the electronic and optical properties of the graphene-based vdW heterostructures constructed by vertical stacking of graphene (G) and CX (X = S, Se and Te) monolayers using first-principles based density functional theory (DFT).

## 2. Computational details

The electronic and optical properties of a CX/G heterostructures have been carried out in Quantum Espresso (QE) simulation package [13] within the framework of DFT. The generalized gradient approximation (GGA) within the Perdew, Burke and Ernzerhof (PBE) scheme is treated as exchange-correlation energy functional in a company with norm-conserving Vanderbilt pseudopotentials have been utilized during structural minimization and other property calculations. For ionic minimization, the chosen convergence criteria for energy and force are $10^{-5}$ a.u. and $10^{-4}$ a.u., respectively. The Brillouin zone with a $12 \times 12 \times 1$ Monkhorst-Pack k-mesh and the plane wave cut-off is $100\ eV$ used for geometry optimization. Further, the selected convergence threshold is $10^{-8}$ a.u. for the electronic property calculations determined by solving the many-body Kohn-Sham equation through the iterative solution method. In 2D semiconductors, the PBE functional underestimate the electronic bandgap value, hence, in this vein, the Heyd-Scuseria-Ernzerhof (HSE06) functional is one of the great choices. However, such functional generally do not change the nature and trend of the band edges and of course demanding high computing cost for large systems, these calculations are not considered for the present work [24,45,50]. The optical properties of CX/G heterostructure were predicted using the real and imaginary parts of complex dielectric function by solving the Kramers-Krönig transformation with a high dense k-mesh of $25 \times 25 \times 1$. A large vacuum space of 20 Å along the z-axis was applied to break the periodicity for 2D nature of the structure. Furthermore, to analyze the influence of biaxial strain ($\varepsilon_{strain}$) on the electronic and optical properties, we have applied both compressive and tensile strains in the range of $(-6\%\ \text{to} +6\%)$ with an 2% interval using the following expression [34]:

$$\varepsilon_{strain} = \frac{a - a^0}{a^0} \times 100\% \qquad (1)$$

Where, $a^0$ and a denotes value of lattice parameter of the unstrained and strained CX/G heterostructures, respectively.

## 3. Results and discussion

### 3.1 Structural properties

We consider planar geometry of CX monolayers and graphene to form 2D vdW heterostructures via vertical staking as depicted in Figure 1. In which the xy plane depicts the top view and yz plane demonstrates the side view of relaxed geometry of considered CX/G heterostructures. The vertical interlayer spacing between the graphene and CX monolayers is set to be 1.5 Å during all properties assessment. The hexagonal lattice of CX/G heterostructures contains three carbons (two of graphene and one from CX monolayer) and one chalcogen atom in the unit cell of the system. The optimized lattice constant for these heterostructures is CS/G (4.056 Å), CSe/G (4.330 Å) and CTe/G (4.600 Å). As noticed, the lattice constant is gradually raised towards the heavier systems due to increasing in the atomic radii of the chalcogen atom in the unit cell. Furthermore, the interlayer binding energy ($E_b$) of CX/G heterostructures is also calculated by $E_b = E_{CX/G} - E_{CX} - E_G$. Here, the $E_{CX/G}$ term denotes the total ground states energy of CX/G heterostructure, whereas, the terms $E_{CX}$ and $E_G$ stand for the total energy of isolated CX and graphene monolayers, respectively. The computed binding energies of CS/G, CSe/G and CTe/G heterostructures are $-4.07, -5.02$ and $-4.64\ eV$, respectively, rendering that the CX/G heterostructures exhibit medium interlayer interactions with stable geometries.

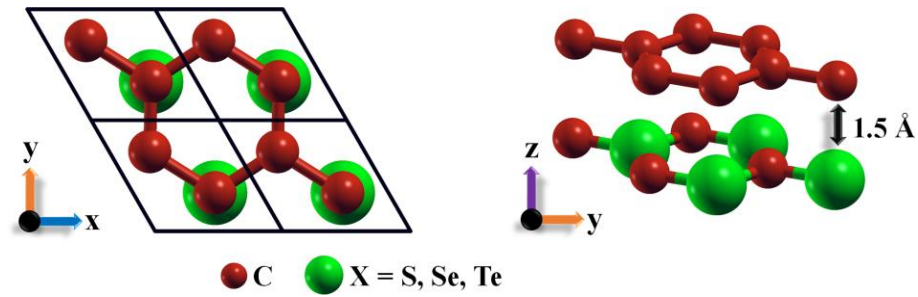

**Figure 1:** Left side represents the top view and right one demonstrates the side view of optimized geometry of CX/G vdW heterostructures. The black box in top view shows the unit cell of hexagonal crystal structure and the vertical interlayer distance between the CX and graphene monolayers is 1.5 Å as shown in the side view diagram.

Strain is considered an essential parameter to influence the physical properties of any material. Therefore, in the present study, we have applied the biaxial strain using equation 1 to investigate the strain impact on the electronic and optical properties of the CX/G heterostructures. The chosen range of strain is from −6% to+6% by varying the optimized lattice constants with the strain interval of the 2%. Here, the '−' sign represents the compressive and '+' sign denotes the tensile strain. Figure 2 picturized the strain-modified lattice constant and variation in pressure in CX/G heterostructures under biaxial strain. It is quite obvious that for compressive strain the lattice constant of heterostructures decreases and increases for the tensile case. Further, during ground state calculations, we also determine the pressure of heterostructures under both strains. It is clearly seen that the pressure is directly proportional to compressive strain while inversely proportional to the tensile strain, further leading to changes in the electronic bandgap value of CX/G heterostructures (See Table 1).

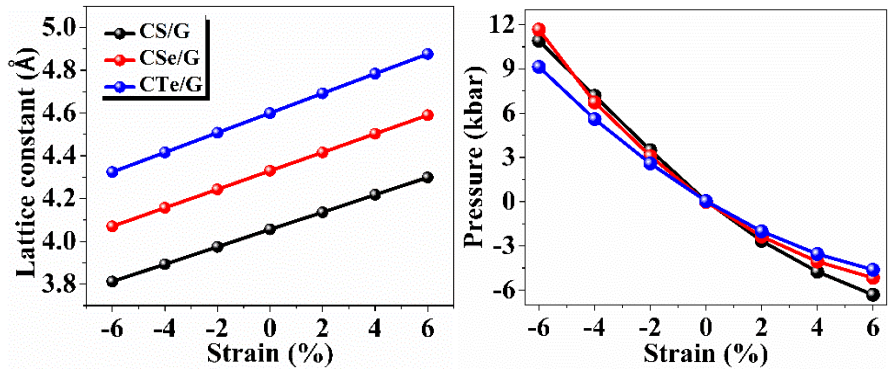

**Figure 2:** Strain induced lattice constant and pressure of CX/G heterostructures.

### 3.2 Electronic properties

#### *3.2.1 Unstrained electronic band structure and effect of SOC*

The electronic properties of CX/G heterostructures are calculated without and with spin orbit coupling (SOC) using the GGA+PBE functional. In Figure 3, the left panel represents the estimated partial density of states (PDOS) of CX/G heterostructures as well as their respective electronic band diagrams at high-symmetry k-path $\mathbf{M} - \mathbf{\Gamma} - \mathbf{K} - \mathbf{M} - \mathbf{\Gamma}$. On the opposite side, the right panel depicts the band edges of valence band maxima (VBM) and conduction band minima (CBM) for both cases (i.e., without and with SOC). According to PBE, all three heterostructures are indirect semiconductors with bandgap values of CS/G ($0.50\ eV$), CSe/G ($0.62\ eV$) and CTe/G ($0.47\ eV$). In the case of PBE+SOC, the obtained bandgap values for these corresponding heterostructures are 0.50, 0.61 and 0.43 $eV$, respectively. It is clearly seen that the predicted bandgap for the PBE+SOC case is quite close to the PBE case, notably for S and Se systems (See Table 1). Moreover, it is evident that the indirect semiconducting nature of CX/G heterostructures is not affected while introducing the SOC. The formation of CX/G heterostructures opens a bandgap of graphene as also evident by literatures [3,34]. In each heterostructure, the VBM is located at **M** point while the CBM is located between $\mathbf{\Gamma - K}$ path. Similarly, for the SOC case, the CBM remains on the same path while the VBM is shifted to the $\mathbf{M - \Gamma}$ path. In the SOC case, the bandgap values for CSe/G and CTe/G heterostructures are reduced due to the higher splitting of bands near the band edges compared to the CS/G system in which the bandgap remains unaffected. The variation in bandgap value for the CTe/G heterostructure is higher than for the other system, which might be due to the presence of a heavier chalcogen atom (i.e., Te) in the system. Furthermore, the obtained bandgap values for these heterostructures are exceedingly low due to the presence of gapless graphene. In other words, one can also imply that these graphene-based heterostructures introduce a significant amount of bandgap, indicating a semiconducting nature with potential applications in optoelectronic devices. Additionally, the valence band edges have a distinct shape known as a "pudding mold" [19], which drives large thermopower of these heterostructures with high Seebeck coefficient and electrical conductivity.

Besides, the strong band splitting was found in the valence and conduction bands of CX/G heterostructures with introducing the SOC, which might be attributed to the lack of an inversion center [21]. Figure 3 shows the higher splitting of bands for the heavier chalcogen CTe/G heterostructure. The calculated band splitting energy (ΔE) at the CBM for CX/G heterostructures followed the trend as: CTe/G (78.4 meV) > CSe/G (24.6 meV) > CS/G (4.4 meV) and summarized in Table 2. While at the VBM, the obtained values of ΔE for the corresponding heterostructures are 17.1, 2.8 and 0.1 meV, respectively. It is clearly seen that the CBM exhibits much stronger band splitting than the VBM, for each CX/G heterostructure. We have found that our obtained ΔE is quite comparable to recently reported theoretical work on the SnS/SnSe based heterostructures for a similar kind of band splitting at the CBM [21]. Along with ΔE, the Table 2 also included the estimated effective masses of electrons $(m_e^*)$ and holes $(m_h^*)$ by parabolic fits from the band diagrams of CX/G heterostructures without and with SOC. It is observed that the electron effective masses steadily increase towards heavier chalcogen heterostructures in both without and with SOC cases, whereas, for the same the hole effective masses for CTe/G heterostructure are smaller compared to the other two heterostructures. Such low hole effective mass indicates the high dispersion nature of the valence band edge of the CTe/G heterostructure. Further, the flatter bands at CBM results in larger effective masses of holes than electrons in CX/G heterostructures.

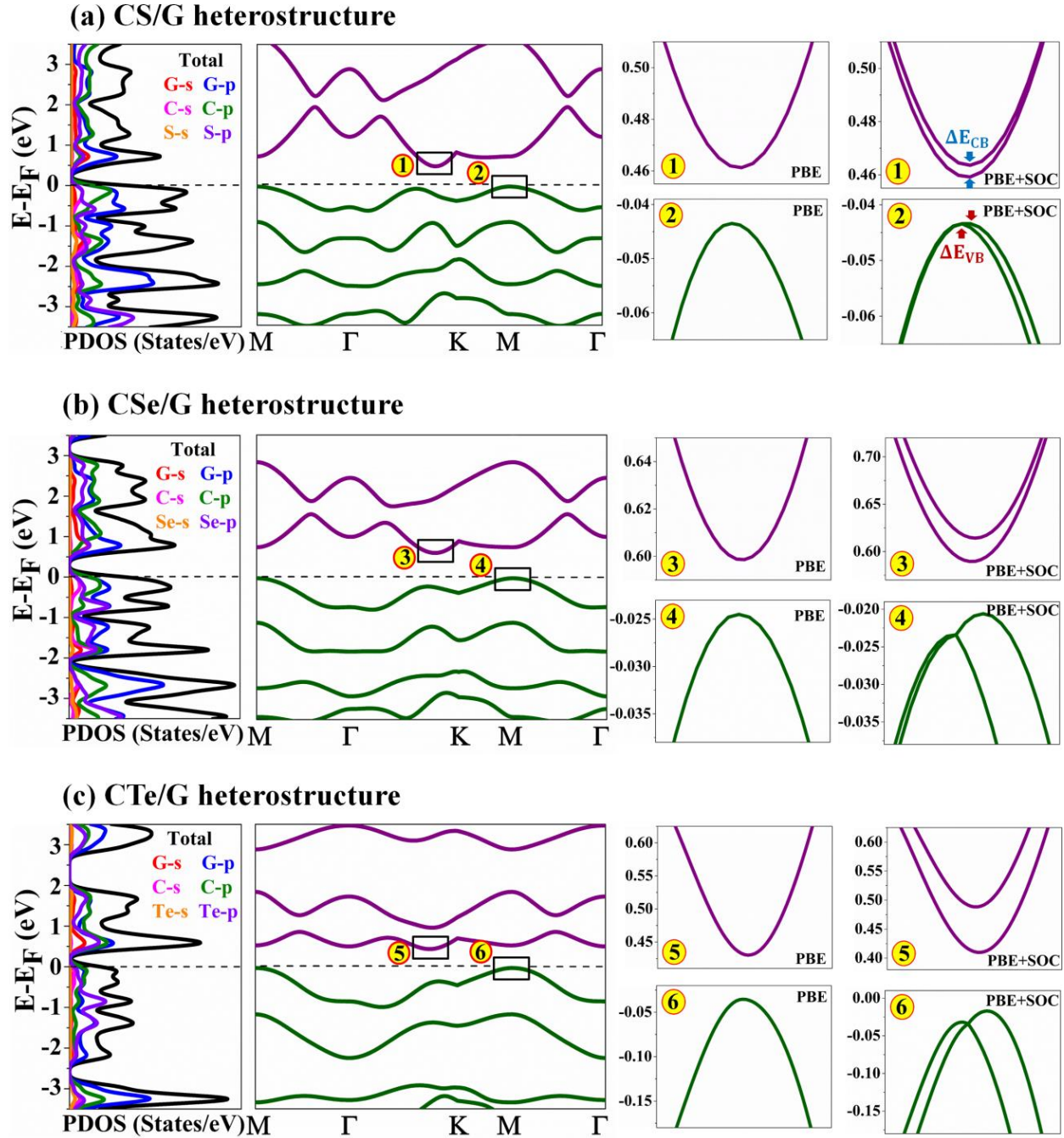

**Figure 3:** Left side panel shows the unstrained electronic band structure and partial density of states (PDOS) of CX/G heterostructures. Whereas, the right side panel demonstrates the magnificent version of valence band maxima (VBM) and conduction band minima (CBM) without and with SOC. The dotted line at zero energy level denotes the Fermi level of the systems.

Moreover, it is noticed that the hole effective masses for CS/G and CSe/G heterostructures are increased while decreasing for CTe/G heterostructure with the incorporation of the SOC calculation and vice versa for electron effective masses. Thus, it can be concluded that the splitting of bands converges the electron and hole effective masses by lowering the degeneracy of the band edges.

**Table 1:** The computed electronic bandgap of unstrained and strained CX/G heterostructures using GGA-PBE functional without and with SOC.

| Strain | CS/G | | CSe/G | | CTe/G | |
|---|---|---|---|---|---|---|
| | $E_g^{PBE}$ (eV) | $E_g^{PBE+SOC}$ (eV) | $E_g^{PBE}$ (eV) | $E_g^{PBE+SOC}$ (eV) | $E_g^{PBE}$ (eV) | $E_g^{PBE+SOC}$ (eV) |
| $-6\%$ | - | - | 0.17 | 0.15 | 0.08 | 0.07 |
| $-4\%$ | 0.16 | 0.16 | 0.33 | 0.32 | 0.18 | 0.16 |
| $-2\%$ | 0.35 | 0.35 | 0.48 | 0.47 | 0.32 | 0.29 |
| $0\%$ | 0.50 | 0.50 | 0.62 | 0.61 | 0.47 | 0.43 |
| $+2\%$ | 0.64 | 0.64 | 0.76 | 0.75 | 0.61 | 0.56 |
| $+4\%$ | 0.76 | 0.76 | 0.90 | 0.87 | 0.76 | 0.69 |
| $+6\%$ | 0.85 | 0.84 | 1.01 | 0.98 | 0.91 | 0.82 |

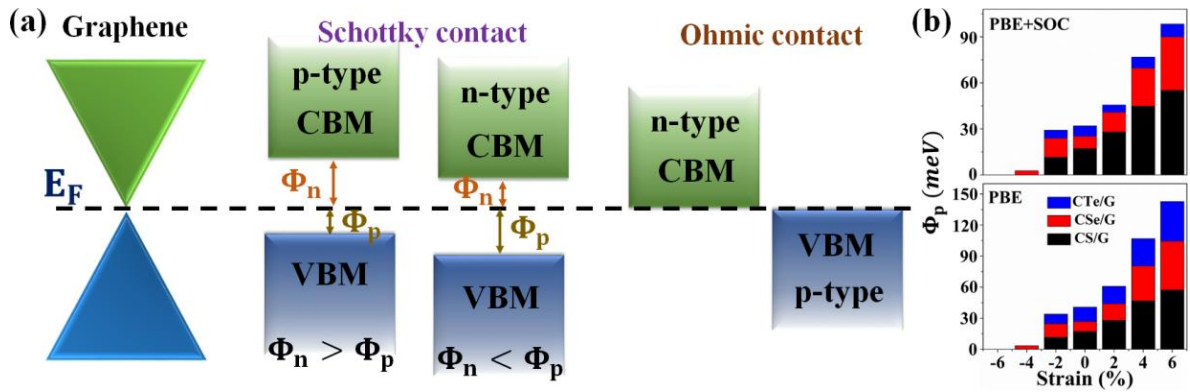

**Figure 4:** (a) Schematic illustration of the Dirac cone of graphene. The energy level of valence and conduction band edges of heterostructures in Schottky contact and Ohmic contact. The dotted line $E_F$ represents the Fermi level. $\Phi_n(\Phi_p)$ denotes the energy difference between the CBM(VBM) and $E_F$. (b) Stacking comparison of $\Phi_p$ of unstrained and strained CX/G heterostructures.

Furthermore, we observed that CX/G heterostructures exhibit a p-type Schottky contact as shown schematically in Figure 4(a). According to the Schottky-Mott rule [40], the Schottky barrier height (SBH) of p- and n-type Schottky contact is calculated by the following expressions:

$$\Phi_{Bp} = E_F - E_{VBM} \qquad (2)$$

$$\Phi_{Bn} = E_{CBM} - E_F \qquad (3)$$

Where, $\Phi_{Bp}$ and $\Phi_{Bn}$ are the SBH of electron and hole of CX/G heterostructures, respectively. The terms $E_{VBM}$, $E_{CBM}$ and $E_F$ denoted the VBM, CBM and Fermi level, respectively. Herein, the calculated $\Phi_{Bp}(\Phi_{Bn})$ of CS/G, CSe/G and CTe/G heterostructures are 43(461), 24(598) and 35(430) $meV$, respectively, for the PBE case. While for the PBE+SOC, the same SBH for the corresponding heterostructures are 43(459), 20(589) and 17(410) $meV$, respectively. The aforementioned SBH values indicate that when SOC is applied, the SBH of VBM and CBM decreases due to band edge splitting; however, SOC computations have no effect on the p-type of Schottky constant and indirect semiconducting nature of CX/G heterostructures.

In addition, the computed PDOS spectra along with the total DOS for the PBE case disclose the contributions of the 's' and 'p' orbitals as well as the bonding characteristics of the graphene and CX systems. As shown in Figure 3, the contribution of 'p' orbitals of carbon atoms is dominating in the total DOS than the other chalcogen atoms 's' and 'p' orbitals. One of the possible reasons for such a higher contribution of the 'p' orbital may be due to the presence of sp$^2$ hybridization in systems. Furthermore, as the system becomes heavier, the contribution of the 'p' orbital of the chalcogen atom gradually increases in both the valence and conduction bands. It should be noted that the total DOS spectra contain several van Hove singularities, rendering the 2D nature of the considered heterostructures [23].

*3.2.2 Strained electronic band structure of heterostructure*

To investigate the influence of biaxial strain on the electronic band diagrams of CX/G heterostructures, an external strain of −6% to +6% with 2% interval was applied to the lattice constant using equation 1. Figure 5 depicts the computed strain-modulated electronic band diagrams of CX/G heterostructures for the GGA+PBE case. It is observed that the bandgap value gradually increases with tensile strain while showing an opposite trend for compressive strain for each heterostructure. Moreover, the incorporation of SOC further reduces the bandgap values for both compressive and tensile strains as listed in Table 1. While moving towards the heavy systems i.e., from S to Se to Te, the reduction in bandgap gradually increases due to strong band splitting in VBM and CBM, rendering that SOC effectively dominates in the heavier system rather than lighter systems. For example, at +6% strain, the reduction in bandgap value in the presence of SOC for CS/G heterostructure is ~0.01 $eV$, but it increases for CSe/G by ~0.03 $eV$ and CTe/G by ~0.09 $eV$. On the other hand, the compressive strain shows a very slight reduction in the bandgap of heterostructures, which further indicates that SOC efficiently dominates with the tensile strain for the present case of heterostructures. However, it may differ for other 2D and bulk materials under various environmental circumstances.

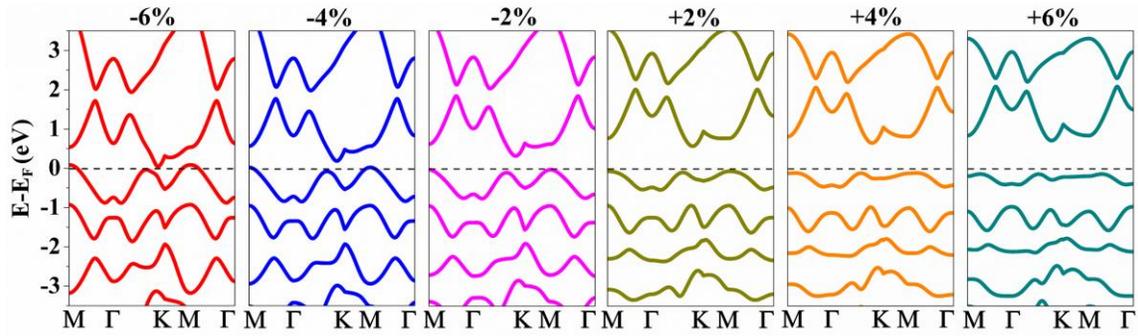

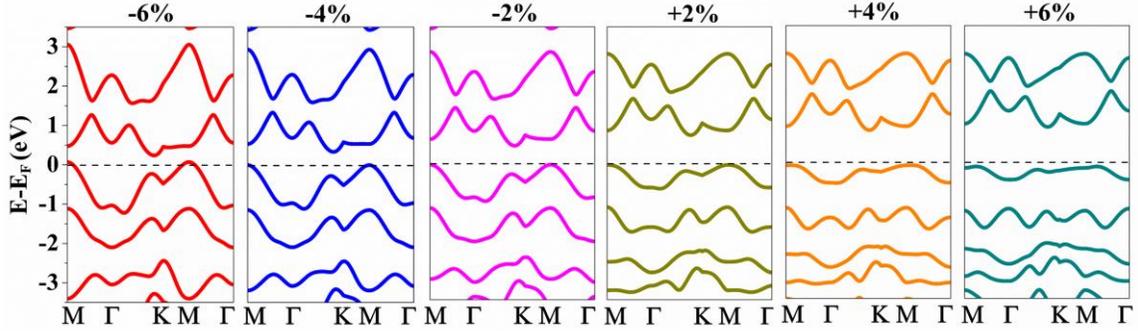

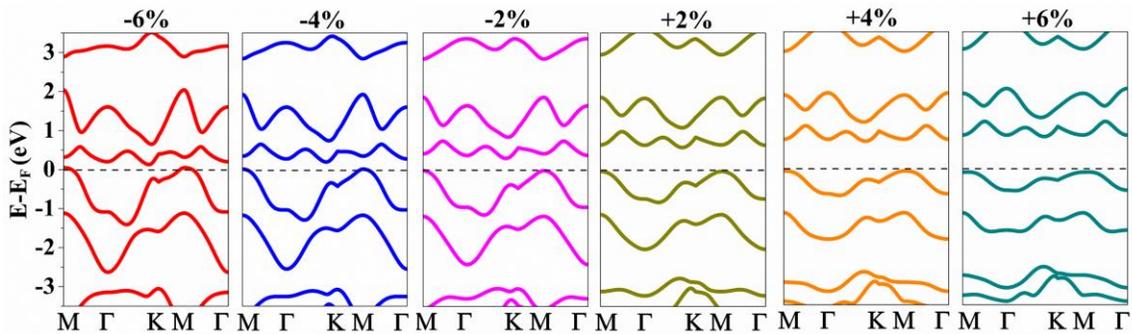

**Figure 5:** The computed electronic band structures of CX/G heterostructures under compressive and tensile strains. The dotted line at zero energy level denotes the Fermi level of the systems.

It is evident that the strain also influences the splitting of valence and conduction band edges as shown in Figure 6. The computed ΔE for each strained heterostructure are listed in Table 2. In CS/G and CSe/G heterostructures, the ΔE for CBM increases with compressive strain while decreasing for the tensile case and vice versa for VBM. Therefore, one can imply that the band splitting at CBM is getting much stronger under compressive strain while weaker for the tensile cases for both CS/G and CSe/G heterostructures, whereas it shows an opposite trend for VBM. On the other hand, the CTe/G heterostructure shows an unpredictable trend as the ΔE for CBM decreases for both compressive and tensile strains, whereas for the case of VBM, the ΔE decreases with compressive strain unto 4%, while increases for 6% and for the tensile case, it continuously keeps increasing. This implies that unstrained CTe/G heterostructure exhibits stronger band splitting at CBM than strained systems and at VBM, stronger band splitting is found for the tensile strain of 6%. Such variation in ΔE of CBM and VBM under compressive and tensile strains is due to the nature of their respective band edges splitting.

Moreover, the calculated effective masses of electrons and holes under compressive and tensile strains are also summarized in Table 2. Similar to unstrained systems, the hole effective masses are larger than the electron effective masses for both compressive and tensile strains owing to flatter bands at CBM. It is observed that the electron effective masses remain almost the same under both compressive and tensile strains attributed to the quite similar nature of CBM. On the counter side, the hole effective masses under compressive strain are nearly constant as compared to the tensile case, which has substantially higher hole effective masses for both without and with SOC. Such large enhancement in effective masses of holes indicates that the bands are getting more flatter at CBM also confirmed by Figure 5&6.

Further, we have found that each heterostructure under tensile strain remains a p-type Schottky contact due to enhancement in the value of $\Phi_{Bp}$ whereas under compressive strain, all three heterostructures are converted into the p-type Ohmic contact because of their VBM overlaps to the Fermi level and thus, $\Phi_{Bp}$ is almost close to zero, as demonstrated in the stacking diagram of Figure 4(b). Hence, it indicating that by using biaxial strain one can easily control the SBH and contact type to tune the electronic properties of these heterostructures.

**(a) CS/G heterostructure**

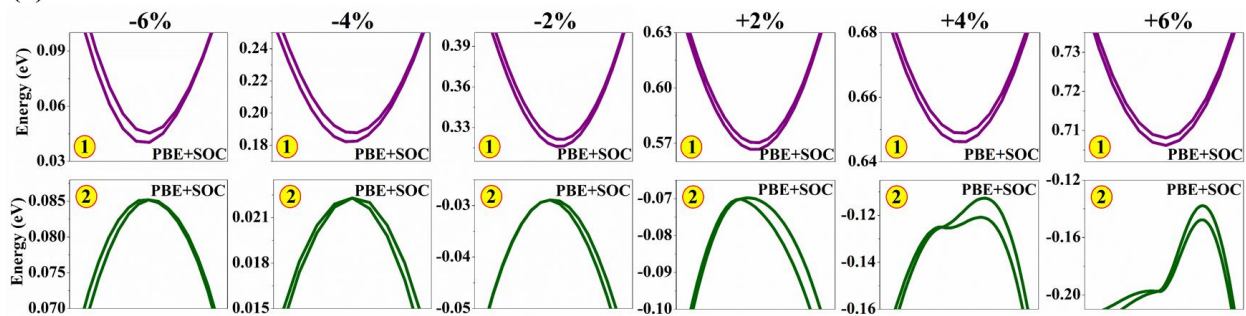

**(b) CSe/G heterostructure**

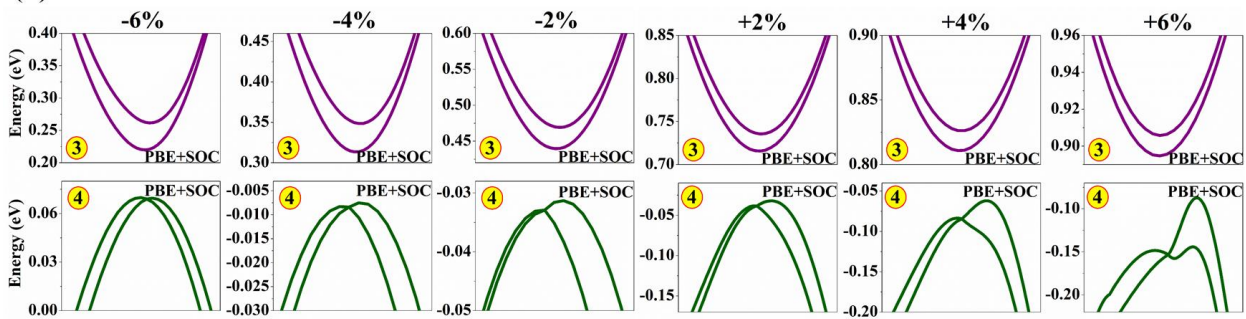

**(c) CTe/G heterostructure**

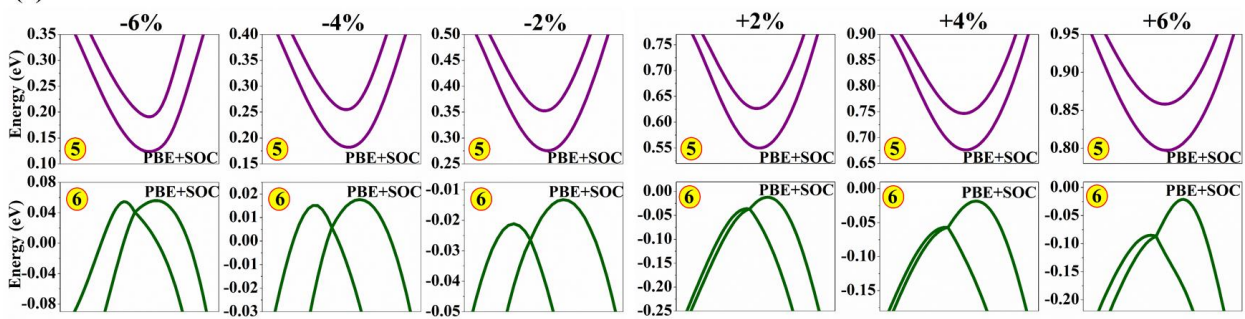

**Figure 6:** The magnificent version of valence band maxima and conduction band minima of CX/G heterostructures under compressive and tensile strains in the presence of SOC.

**Table 2**: The computed band splitting in conduction ($\Delta E_{CB}$) and valence ($\Delta E_{VB}$) bands and their respective effective masses in the unit of $m_0$ of unstrained and strained CX/G heterostructures in the presence of SOC. Note: The calculated effective masses for PBE case is written in the bracket along with the effective masses of PBE+SOC values.

| Parameters | Strain | | | | | | |
|---|---|---|---|---|---|---|---|
| | $-6\%$ | $-4\%$ | $-2\%$ | $0\%$ | $+2\%$ | $+4\%$ | $+6\%$ |
| **CS/G heterostructure** | | | | | | | |
| $\Delta E_{CB}$ (meV) | - | 5.5 | 5.2 | 4.4 | 3.6 | 2.7 | 1.9 |
| $\Delta E_{VB}$ (meV) | - | 0 | 0 | 0.1 | 0.3 | 8.2 | 9.9 |
| $m_e^*$ ($m_0$) | - | 0.019 (0.015) | 0.019 (0.020) | 0.021 (0.022) | 0.022 (0.023) | 0.0023 (0.023) | 0.024 (0.023) |
| $m_h^*$ ($m_0$) | - | 0.050 (0.048) | 0.053 (0.051) | 0.074 (0.070) | 0.168 (0.131) | 0.186 (0.164) | 0.100 (0.100) |
| **CSe/G heterostructure** | | | | | | | |
| $\Delta E_{CB}$ (meV) | 41.8 | 35 | 29.6 | 24.6 | 19.9 | 15.5 | 11.1 |
| $\Delta E_{VB}$ (meV) | 0.9 | 0.9 | 1.6 | 2.8 | 6.5 | 22.5 | 56.9 |
| $m_e^*$ ($m_0$) | 0.032 (0.033) | 0.033 (0.035) | 0.034 (0.034) | 0.033 (0.034) | 0.033 (0.033) | 0.033 (0.033) | 0.033 (0.032) |
| $m_h^*$ ($m_0$) | 0.048 (0.039) | 0.053 (0.045) | 0.069 (0.055) | 0.078 (0.077) | 0.131 (0.131) | 0.124 (0.253) | 0.101 (0.110) |
| **CTe/G heterostructure** | | | | | | | |
| $\Delta E_{CB}$ (meV) | 67.1 | 72.8 | 77.1 | 78.4 | 76.5 | 70.7 | 60.8 |
| $\Delta E_{VB}$ (meV) | 25.3 | 11.9 | 13.2 | 17.1 | 24.6 | 38.9 | 65.7 |
| $m_e^*$ ($m_0$) | 0.043 (0.032) | 0.042 (0.034) | 0.044 (0.038) | 0.048 (0.045) | 0.053 (0.051) | 0.056 (0.056) | 0.058 (0.057) |
| $m_h^*$ ($m_0$) | 0.084 (0.048) | 0.064 (0.048) | 0.066 (0.057) | 0.071 (0.073) | 0.083 (0.095) | 0.089 (0.184) | 0.097 (0.298) |

## 3.3 Optical properties and solar parameters

This section will provide a detailed discussion of the optical properties of both unstrained and strained CX/G heterostructures and the impact of biaxial strain on solar power conversion efficiency. It is well-known fact that the generation of electrical energy from sunlight is directly related to the optical performance of the material for their solar energy applications. In this manner, the understanding of optical properties is a vital requirement in order to determine the solar performance of the material. Further, it is also significant to note that the interaction mechanism between light with electrons and ions also depends on the optical performance of the material. Besides, the optical properties and solar performance of the material can also be improved under the influence of biaxial strain [26,39]. Therefore, the optical properties of unstrained and strained CX/G heterostructures were first calculated by computing the complex dielectric function $\varepsilon(\omega)$, which is an addition of two parts: real $\varepsilon_1(\omega)$ and imaginary $\varepsilon_2(\omega)$. Here, the $\varepsilon_1(\omega)$ and $\varepsilon_2(\omega)$ have been computed using the solution of Kramers-Krönig equations [4]. Figure 7 represents the computed $\varepsilon_1(\omega)$ and $\varepsilon_2(\omega)$ spectrums as a function of the energy of CX/G heterostructures without and with strain. The $\varepsilon_1(\omega)$ spectra demonstrate the non-linear optical properties of the material whereas the light absorption characteristics are determined from the $\varepsilon_2(\omega)$ spectra. The picturized $\varepsilon_1(\omega)$ plots of all three heterostructures in Figure 7 demonstrate the ability of polarization of light of CX/G heterostructures and at zero frequency it gives the value of a static dielectric constant, depicted in Table 3. It is clearly seen that the static dielectric constant gradually increases as the strain varies from compressive ($-6\%$) to tensile ($+6\%$) for each heterostructure. Further, a similar trend has been observed towards the heavier chalcogen heterostructure, i.e., the static dielectric constant for each unstrained and strained heterostructure is CTe/G > CSe/G > CS/G, as shown in Table 3. Such enhancement in the value of a static dielectric constant in heavier systems confirms that the electrons of valence bands are more active than those in core bands. Moreover, it also indicates the existence of excitons, which are a pair of mildly bound electrons and holes that are responsible for the better optoelectronic applications of the material through evaluating the carrier lifetime and recombination rates with high optical polarization [14]. The $\varepsilon_1(\omega)$ spectra of CX/G heterostructures exhibit negative values between the energy range of 1-5 $eV$ for compressive strains of $-4\%$ and $-6\%$, It implies that both of the heterostructures will act as metallic systems for this energy region [9]. Whereas, in the other cases of compressive (i.e., $-2\%$), all tensile and unstrained systems, the $\varepsilon_1(\omega)$ spectra remain positive throughout the energy region, rendering the semiconducting nature of the CX/G heterostructures. The computed $\varepsilon_2(\omega)$ plots (See Figure 7) basically explains the light absorption characteristics of the CX/G heterostructures, which have a very close relationship with the electronic properties. As shown in Figure 7, the $\varepsilon_2(\omega)$ spectrum in each unstrained and strained system achieves a non-zero value in the energy region that is almost close to the calculated electronic bandgap values, confirming that all the considered heterostructures are narrow-gap indirect semiconductors. Further, the $\varepsilon_2(\omega)$ spectra aid in understanding the radiative photon emission phenomenon that occurs inside the material during momentum transfer [17]. Furthermore, the nature of $\varepsilon_2(\omega)$ spectra in each unstrained and strained heterostructure reveals the transparent behaviours of the systems due to complete reflections of the incident photons in the low-energy region.

After carefully evaluating the $\varepsilon_1(\omega)$ and $\varepsilon_2(\omega)$ parts of $\varepsilon(\omega)$, we further investigated the optical absorption coefficient ($\alpha$) as a function of energy for each unstrained and strained CX/G heterostructure (See Figure 7). Note that knowledge of $\alpha$ is an essential requirement in order to understand the optical performance of the material [1,17] and it is also utilized to determine the optical bandgap ($E_{og}$) of the material. The interpretation of $\alpha$ discloses the light absorption capability of the material and it also gives the amount of a particular wavelength of light that will be penetrated inside the material. If the absorption peak of any material falls within the optical spectra region, it can be used in optoelectronic and photovoltaic devices. Figure 7 shows the absorption peak of each unstrained CX/G heterostructure falls in the visible spectra, rendering the better light absorption ability for solar energy conversion devices. However, in the case of strain, the absorption peak shifts towards the low-energy region (i.e., blue shift) for the compressive strain whereas for tensile strain, it moves to the high-energy region (i.e., red shift). By utilizing $\alpha$, the $E_{og}$ of CX/G heterostructures have been calculated via the Tauc plot and summarized in Table 3. It is evident that the $E_{og}$ varies with applied strain in the same way that the electronic bandgap and static dielectric constant do, i.e., they are all inversely proportional to compressive strain and directly proportional to tensile strain. Moreover, the $E_{og}$ for both the unstrained and strained CTe/G heterostructures is comparatively much lower than the other two systems, suggesting their far better solar energy conversion abilities.

Afterward, the solar power conversion efficiency ($\eta$) of both unstrained and strained CX/G heterostructures have been calculated by the following expression:

$$\eta = \frac{J_{sc} \times V_{oc} \times FF}{P_{in}} \quad (4)$$

Where the short circuit current density defines as $J_{sc}$ and calculated using the available data of air mass 1.5 global (AM1.5G) solar spectrum using equation (5) and the input power $P_{in} = 1000 \; W/m^2$.

$$J_{sc} = q \int_{E_g}^{\infty} \frac{S(E)}{E} dE \qquad (5)$$

The term $S(E)$ denotes the power of incident photons on the earth per unit area with energy $E$. The computed $J_{sc}$ was further used to plot the J-V curve for calculating the open-circuit voltage ($V_{oc}$). Moreover, the Fill factor is calculated by, $FF = \frac{\vartheta_{oc} - \ln(\vartheta_{oc} + 0.72)}{\vartheta_{oc} + 1}$; Where the $\vartheta_{oc} = \frac{q \times V_{oc}}{k_B T}$; $q$ and $k_B$ represent the elementary charge and Boltzmann constant, respectively. Table 3 summarized the calculated $J_{sc}$, $V_{oc}$, $FF$ and $\eta$ of each unstrained and strained CX/G heterostructure. It is observed that $V_{oc}$ and $FF$ both follow the same trend like $E_{og}$ and $E_g^{PBE}$ whereas, the $J_{sc}$ and $\eta$ shows opposite trend with respect to strain. This indicates that $J_{sc}$ is inversely while $V_{oc}$ is directly proportional to $E_{og}$. The optimal value of all these three parameters are highly essential in order to achieve a much better $\eta$ for solar cell applications [1,17]. From Table 3, it is found that the CTe/G heterostructure possess exceptionally high $J_{sc}$ and low $V_{oc}$, which in result achieves excellent $\eta$ as compared to the other two systems.

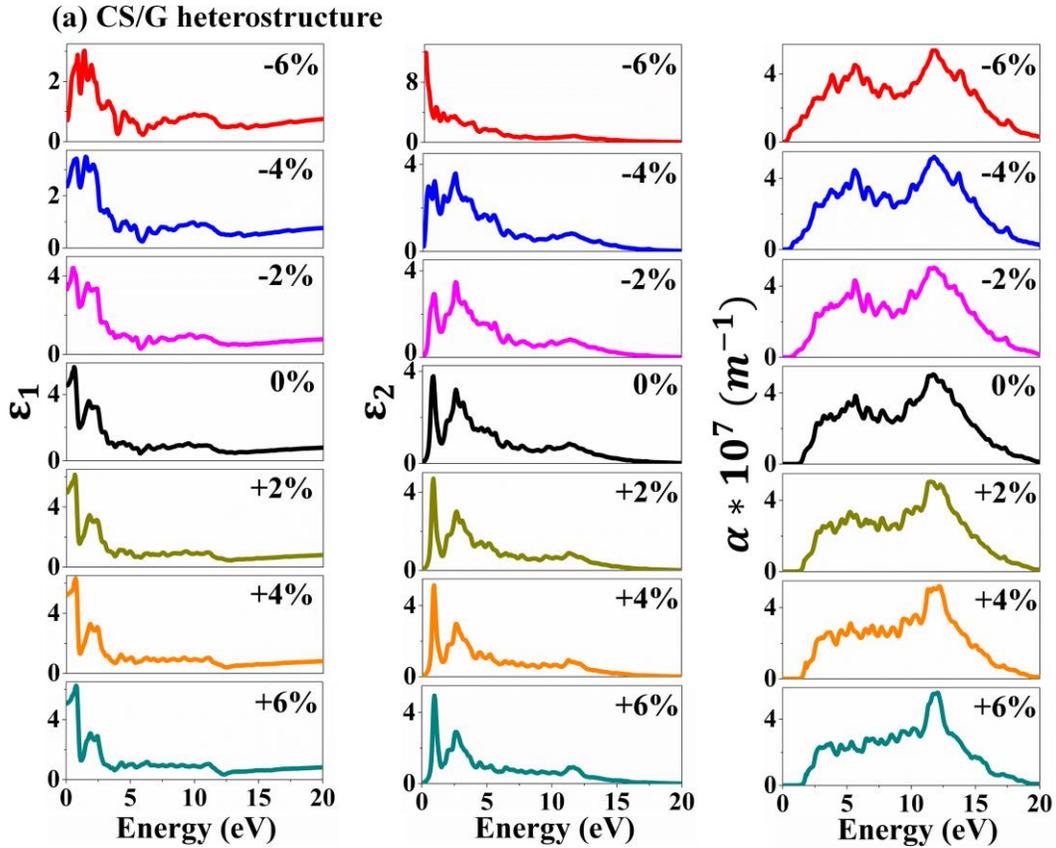

(a) CS/G heterostructure

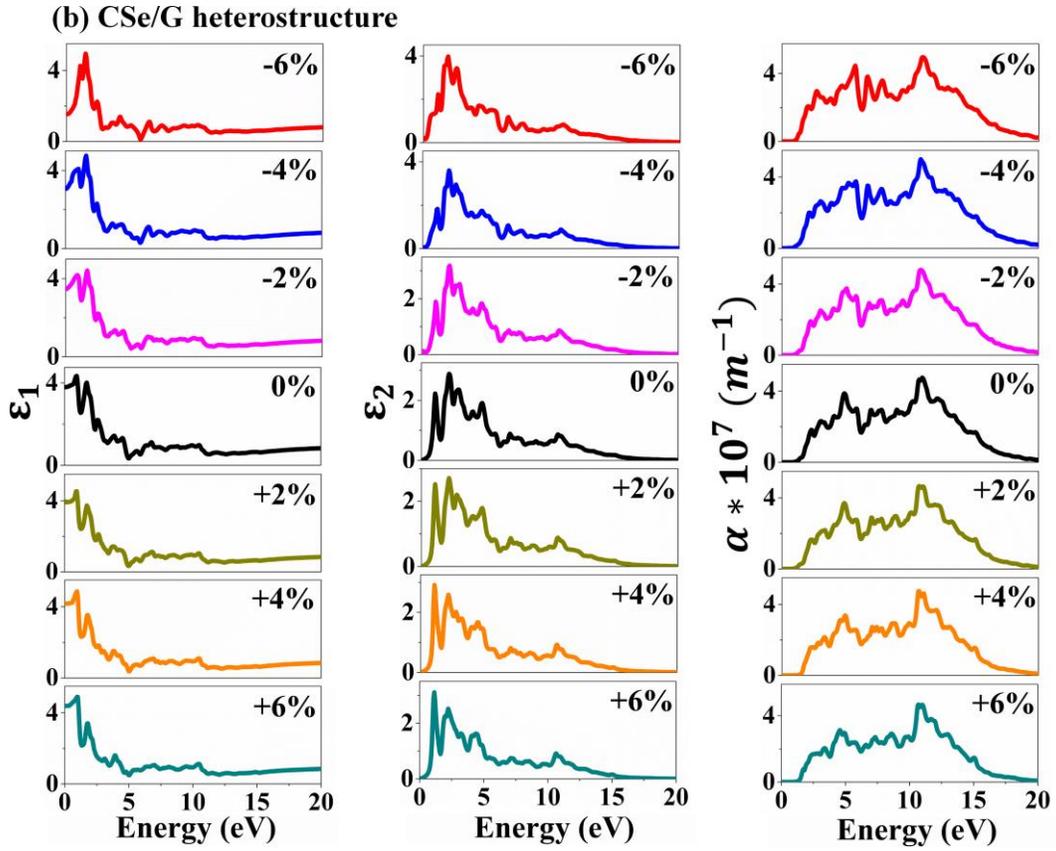
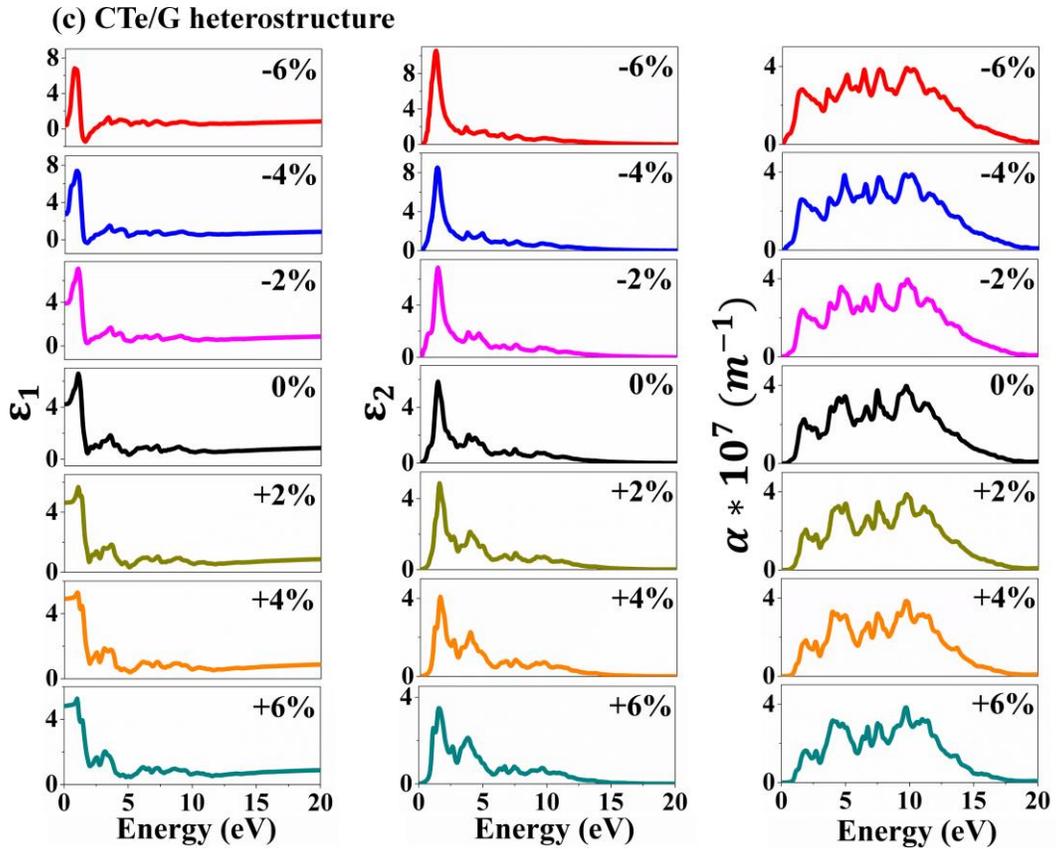

**Figure 7:** The computed optical parameters: real part ($\varepsilon_1$), imaginary part ($\varepsilon_2$) and absorption coefficient ($\alpha$) of unstrained and strained CX/G heterostructures as a function of energy (eV).

**Table 3:** The computed static dielectric constant, optical bandgap ($E_{og}$), short-circuit current density ($J_{sc}$), open-circuit voltage ($V_{oc}$), fill factor ($FF$) and solar power conversion efficiency ($\eta$) of unstrained and strained CX/G heterostructures.

| Parameters | Strain | | | | | | |
|---|---|---|---|---|---|---|---|
| | **−6%** | **−4%** | **−2%** | **0%** | **+2%** | **+4%** | **+6%** |
| **CS/G heterostructure** | | | | | | | |
| Static dielectric constant | 0.71 | 2.36 | 3.33 | 4.53 | 4.96 | 5.06 | 5.21 |
| $E_{og}$ (eV) | - | 2.76 | 3.13 | 3.36 | 3.48 | 3.72 | 4.08 |
| $J_{sc}$ (A/m$^2$) | - | 34.13 | 12.45 | 6.87 | 4.66 | 1.56 | 0.0045 |
| $V_{oc}$ (V) | - | 2.40 | 2.74 | 2.95 | 3.06 | 3.27 | 3.47 |
| $FF$ (%) | - | 94.20 | 94.79 | 95.09 | 95.24 | 95.49 | 95.71 |
| $\eta$ (%) | - | 7.71 | 3.23 | 1.92 | 1.35 | 0.48 | 0.00 |
| **CSe/G heterostructure** | | | | | | | |
| Static dielectric constant | 1.53 | 3.06 | 3.47 | 3.79 | 3.91 | 4.17 | 4.39 |
| $E_{og}$ (eV) | 2.40 | 2.76 | 2.88 | 3.12 | 3.25 | 3.36 | 3.43 |
| $J_{sc}$ (A/m$^2$) | 75.12 | 34.18 | 25.37 | 12.63 | 9.48 | 6.87 | 5.45 |
| $V_{oc}$ (V) | 2.07 | 2.41 | 2.52 | 2.74 | 2.85 | 2.95 | 3.02 |
| $FF$ (%) | 93.46 | 94.22 | 94.42 | 94.79 | 94.95 | 95.09 | 95.19 |
| $\eta$ (%) | 14.53 | 7.76 | 6.03 | 3.28 | 2.56 | 1.92 | 1.57 |
| **CTe/G heterostructure** | | | | | | | |
| Static dielectric constant | 0.43 | 2.74 | 3.91 | 4.23 | 4.63 | 4.83 | 4.93 |
| $E_{og}$ (eV) | - | 1.68 | 1.80 | 1.92 | 2.04 | 2.28 | 2.42 |
| $J_{sc}$ (A/m$^2$) | - | 222.02 | 196.18 | 164.53 | 136.90 | 92.56 | 72.04 |
| $V_{oc}$ (V) | - | 1.40 | 1.51 | 1.62 | 1.74 | 1.96 | 2.09 |
| $FF$ (%) | - | 91.08 | 91.59 | 92.05 | 92.49 | 93.17 | 93.51 |
| $\eta$ (%) | - | 28.31 | 27.13 | 24.53 | 22.03 | 16.90 | 14.08 |

By utilizing these solar parameters in equation 2, the achieved highest $\eta$ for CTe/G heterostructure is 28.31% for $-4\%$ strain whereas, for the unstained system it is 24.53% and it decreases to 14.08% for $+6\%$ strain. On the other hand, the highest achieved $\eta$ for CSe/G heterostructure is 14.53% and for the remaining one it lies in between the 0-8%. Comparing with our previous reports and other pieces of literature [1,2,15,17,29,38,48], it is found that the achieved $\eta$ for CTe/G heterostructure is much higher due to outstandingly large $J_{sc}$ and much lower value of $V_{oc}$. Moreover, the computed $FF$ of each unstrained and strained CX/G heterostructure is greater than 91.08%, demonstrating the quality of each system. Therefore, the strain-influenced optical properties and the computed solar parameters indicate that CTe/G heterostructure displaying strong absorption in the visible region, making it a suitable candidate for solar cell device applications.

## 4. Conclusions

In summary, we have investigated the strain-modulated electronic and optical properties of CX/G heterostructures based on DFT calculations. According to the computed structural properties and binding energy, all three heterostructures have stable hexagonal crystal geometry. Further, they all shows semiconducting characteristics with indirect bandgap: CS/G (0.50 $eV$), CSe/G (0.62 $eV$) and CTe/G (0.47 $eV$). The incorporation of SOC with electronic property calculations results in the splitting of conduction and valence band edges and a reduction in the bandgap values in CSe/G (0.61 $eV$) and CTe/G (0.43 $eV$) heterostructures. Under biaxial strain, the electronic bandgap for without and with SOC structures further decreases with compressive strain while increasing for tensile case, similar kinds of observations have also been perceived in the band splitting energy and effective masses. The Schottky barrier heights demonstrate that the CX/G heterostructures are p-type Schottky contacts which can be efficiently tuned into p-type Ohmic contacts by applying biaxial strains. Furthermore, the computed real and imaginary parts of the dielectric function reveal the existence of electron-ion pairs, narrow gap semiconducting nature and transparent behaviours of CX/G heterostructures for improved optoelectronic applications. Besides, the strain-tuned optical absorption and computed solar parameters represent the CTe/G heterostructure as the most promising candidate for solar cell devices with an excellent $\eta$ of 24.53% for unstrained structure, rather than the other two heterostructures. We believe that the findings of the present study can potentially guide the researchers to design novel optoelectronic devices based on CX/G heterostructures.

**Acknowledgments**

A.K.B. acknowledge the Government of Gujarat, India for the ScHeme Of Developing High quality research (SHODH) for providing financial support (Ref No. 202001450003). All the computational calculations have been carried out using the workstation located at IITRAM (under the SERB-DST project: ECR/2016/001289).